\documentstyle[11pt,moriond,epsfig]{article}

\def\gev{GeV}                           
\def\etal{{\sl et al.}}                 
\def\Journal#1#2#3#4{{#1} {\bf #2}, #3 (#4)}

\def\NIMA{{\em Nucl. Instrum. Methods} A}
\def\NPB{{\em Nucl. Phys.} B}
\def\PLB{{\em Phys. Lett.}  B}
\def\PRL{\em Phys. Rev. Lett.}
\def\PRD{{\em Phys. Rev.} D}

\def\be{\begin{equation}}
\def\ee{\end{equation}}
\def\bea{\begin{eqnarray}}
\def\eea{\end{eqnarray}}

\begin{document}
\vspace*{4cm} \title{PRELIMINARY MEASUREMENT OF THE DIFFERENTIAL CROSS
  SECTION FROM NEUTRINO-NUCLEON DEEPLY INELASTIC SCATTERING AT NUTEV
}

\author{J.~McDonald$^7$, D.~Naples$^7$, T. Adams$^4$, A.~Alton$^4$,
  S.~Avvakumov$^8$, L.~de~Barbaro$^5$, P.~de~Barbaro$^8$,
  R.~H.~Bernstein$^3$, A.~Bodek$^8$, T.~Bolton$^4$, J.~Brau$^6$,
  D.~Buchholz$^5$, H.~Budd$^8$, L.~Bugel$^3$, J.~Conrad$^1$,
  R.~B.~Drucker$^6$, B.~T. Fleming$^1$, J.~Formaggio$^1$, R.~Frey$^6$,
  J.~Goldman$^4$, M.~Goncharov$^4$, D.~A.~Harris$^8$, J.~H.~Kim$^1$,
  S.~Koutsoliotas$^1$, R.~A.~Johnson$^2$, M.~J.~Lamm$^3$,
  W.~Marsh$^3$, D.~Mason$^6$, K.~S.~McFarland$^8$, C.~McNulty$^1$,
  P.~Nienaber$^3$, V.~Radescu$^7$, A.~Romosan$^1$,
  W.~K.~Sakumoto$^8$,H.~Schellman$^5$, M.~H.~Shaevitz$^1$,
  P.~Spentzouris$^1$, E.~G.~Stern$^1$, N.~Suwonjandee$^2$,
  N.~Tobien$^3$, M.~Tzanov$^7$, A.~Vaitaitis$^1$, M.~Vakili$^2$,
  U.~K.~Yang$^8$, J.~Yu$^3$, G.~P.~Zeller$^5$,
  E.~D.~Zimmerman$^1$\\
  {\it The NuTeV Collaboration}}

\address{ \it \baselineskip=13pt $^1$Columbia University, New York,
  NY, $^2$University of Cincinnati, Cincinnati, OH, $^3$Fermi National
  Accelerator Laboratory, Batavia, IL, $^4$Kansas State University,
  Manhattan, KS, $^5$Northwestern University, Evanston, IL,
  $^6$University of Oregon, Eugene, OR, $^7$University of Pittsburgh,
  Pittsburgh, PA, $^8$University of Rochester, Rochester, NY.}

\maketitle \abstracts{ Preliminary results for the neutrino-nucleon
  differential cross section from the NuTeV experiment are presented.
  The extraction of the differential cross section from NuTeV is
  discussed and the structure functions $F_2$ and $\Delta xF_3$ are
  presented.  Comparisons are made with CCFR results.}

\section*{Introduction}
Neutrino-nucleon experiments offer a rich source of information about
the quark structure of the proton.~\cite{nuRev} Neutrino-nucleon
deeply inelastic scattering (DIS) is arguably the most direct
measurement of the proton structure functions.

The NuTeV experiment observed $6\times10^5$ neutrino ($\nu_{\mu}$) and
$3\times10^5$ anti-neutrino ($\overline{\nu}_{\mu}$) charged-current
interactions in an iron/scintillator calorimeter.~\cite{nutCal} The
NuTeV beam was generated from 800 \gev\ protons on a beryllium oxide
target; secondary pions and kaons were sign-selected using a quadrupole
magnet train.  NuTeV included a precision calibration beam designed to
reduce the uncertainty of the absolute muon and hadron energy scale.

\section*{Extracting the Differential Cross Sections}

The differential cross section is determined from the differential number
of events $\frac{d^2 N}{dxdy}$ and a flux factor $L(E)$ at a given
neutrino energy,
\begin{equation}
  \frac{d^2\sigma^{\nu(\overline{\nu})}}{dxdy} =  \frac{1}{L(E)}\frac{d^2 N^{\nu(\overline{\nu})}}{dxdy}. 
\label{eq:dfe1}
\end{equation}

In Eq.~\ref{eq:dfe1}, the kinematic variables $x$ and $y$ represent
the Bjorken scaling variable (the fractional momentum of the struck
quark), and the inelasticity.  NuTeV reconstructs these kinematic
variables and the energy of the neutrino.  The NuTeV kinematic range
extends from about $10^{-3}$ to 0.95 in $x$ and from 0.05 to 0.95 in
$y$; the energy reach is from 30 GeV to about 400 GeV.

The differential number of events is determined from a sample of
events which pass charged current quality cuts, which demand event
containment, a minimum hadronic energy ($\nu$) of 10~GeV, a momentum
analyzed muon, and a minimum $Q^2$ cut.  The selected events are
binned in $x$, $y$, and $E$; are corrected for detector effects; and
are acceptance corrected using a fast detector simulation.  The
binning is chosen to approximately reflect the detector resolution.

A nearly orthogonal sample of events is selected to determine the
flux.  These events must pass a maximum hadronic energy cut of 20 GeV.
The flux is solely a function of the neutrino energy and is determined
by expanding the flux differential cross section in terms of
$\frac{\nu}{E}$.  To first order, up to corrections of order
$\frac{\nu}{E}$, the flux is proportional to the number of events in
this sample.  The $\frac{\nu}{E}$ corrections are obtained by fitting
for the coefficients of $\frac{\nu}{E}$ and relating the coefficients
to physical quantities.  This procedure determines the relative flux
as a function of energy.  The flux is normalized to the world total
neutrino cross section by using a third sample which includes both the
flux and the charge current sample.

\begin{figure}
\psfig{file= 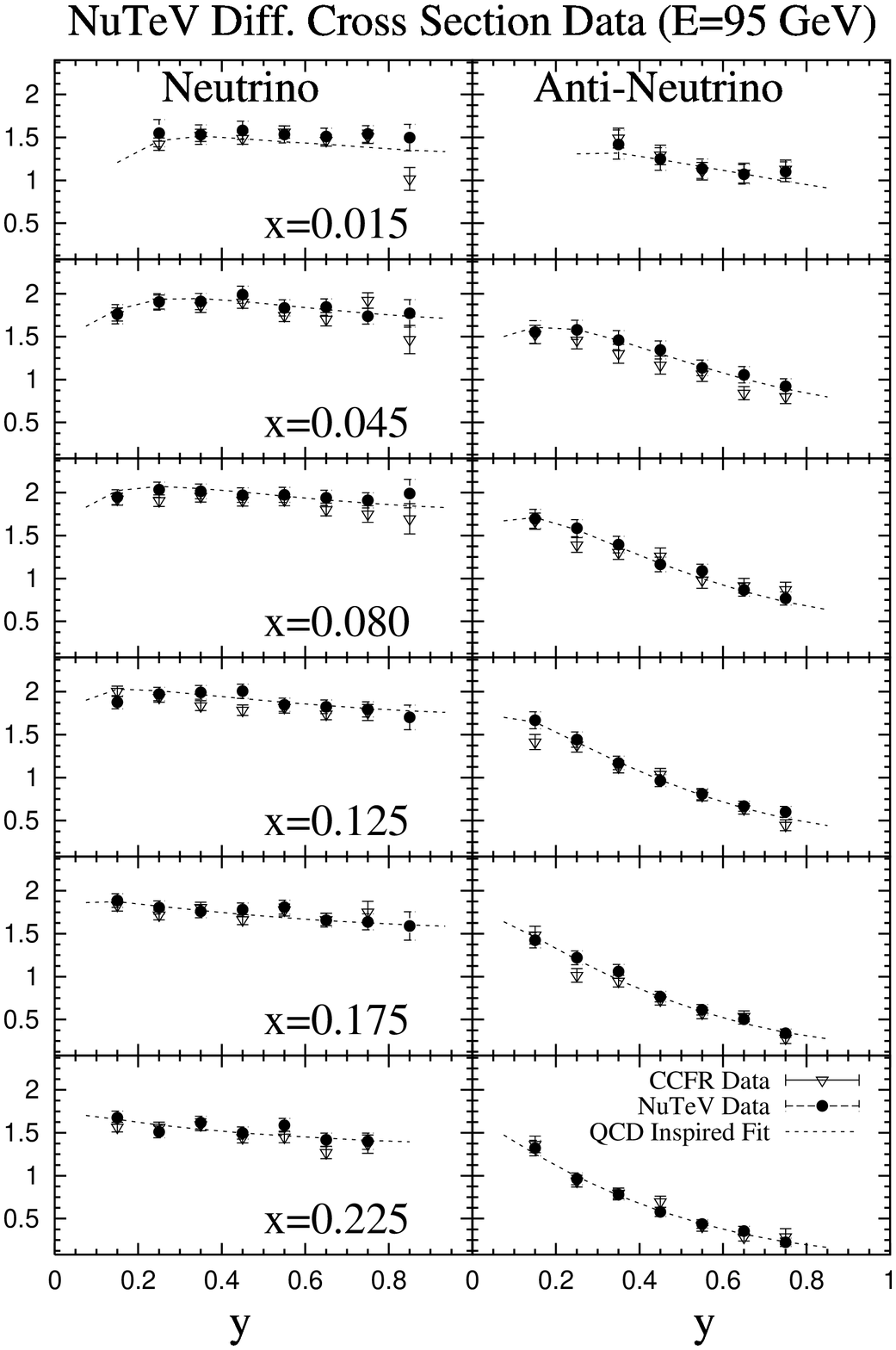, height=9cm}
\psfig{file= 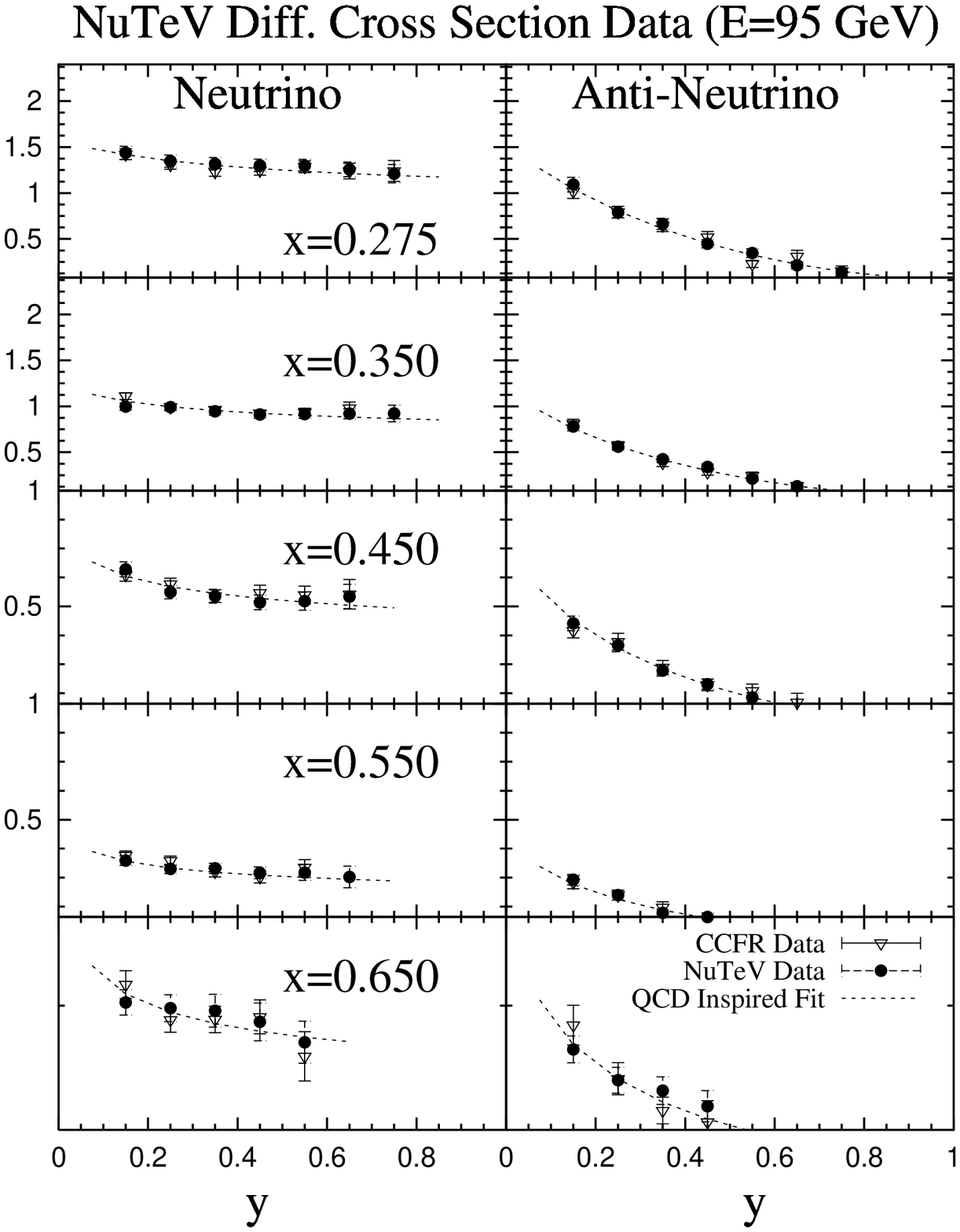, height=9cm}
\caption{Neutrino and anti-neutrino (side by side) differential
  cross section from NuTeV (closed circles) and CCFR (open triangles)
  as a function of $y$ with the QCD inspired fit to the NuTeV data.
  Only statistical errors are shown.  The differential cross section
  should be scaled by $G_F^2 M/E_{\nu}$ to obtain the proper units.
  The dashed line is the NuTeV differential cross section calculation
  using empirically determined parton distribution functions.}
\label{fig:dfxs}
\end{figure}

The fast detector simulation, which takes into account acceptance and
resolution effects, includes an empirically determined set of parton
distribution functions with QCD evolution.~\cite{bg} The parton
distribution functions are determined by fitting the extracted
differential cross section.  The fitted parton distributions are used
to determine acceptance corrections for both the DIS and flux sample.
The procedure is then iterated.  As a typical example, the
differential cross section at $E = 95$ GeV is shown in
Fig.~\ref{fig:dfxs}.  NuTeV is found to be in good agreement with
CCFR.~\cite{ccfr}

%

In terms of $\epsilon$ (the polarization of the virtual $W$) the sum
of the anti-neutrino and neutrino differential cross sections can be
written

\begin{equation}
   F(\epsilon) =  \frac{\pi (1-\epsilon)}{y^2 G_F^2 M_p E_{\nu}}
      \left( \frac{d^2\sigma^\nu}{dxdy} +
        \frac{d^2\sigma^{\overline{\nu}}}{dxdy} \right) 
 =  2xF_1\left[1+\epsilon R(x,Q^2)\right] + 
  {g(y)} \Delta xF_3.
\label{eq:feps2} 
\end{equation}
The rightmost equation allows the function $F(\epsilon)$ to be related
to the underlying physical quantities.  The structure functions
$2xF_1$ and $\Delta xF_3 = x(F_3^{\nu} - F_3^{\overline{\nu}})$ are functions
of $x$ and $Q^2$ only.  The $y$-dependence of the differential cross
section is contained in $\epsilon = \frac{2(1-y)- M_pxy/E}{1+(1-y)^2 +
  M_pxy/E}$ and $ g(y) = \frac{y(1-y/2)}{1+(1-y)^2}$.  The
coefficients from the fit of $F(\epsilon)$ to the terms $(1+\epsilon
R(x,Q^2))$ and $g(y)$ represent $2xF_1$ and $\Delta xF_3$,
respectively.  The structure function $F_2$ is given by $F_2 =
2xF_1(1+R(x,Q^2))$.

\section*{Results}

While the differential cross section offers the most fundamental
picture of neutrino (anti-neutrino) DIS, the structure functions
illuminate the underlying physical quantities.  Unfortunately,
knowledge of both $R(x,Q^2)$ and $\Delta xF_3$ is limited in the
low-$x$, low-$Q^2$ region.~\cite{ccfr,nnlo} $F_2$, $R$, and $\Delta
xF_3$ cannot be simultaneously fit due to inadequate statistics and
strong correlations among the parameters.

\begin{figure}
\psfig{figure=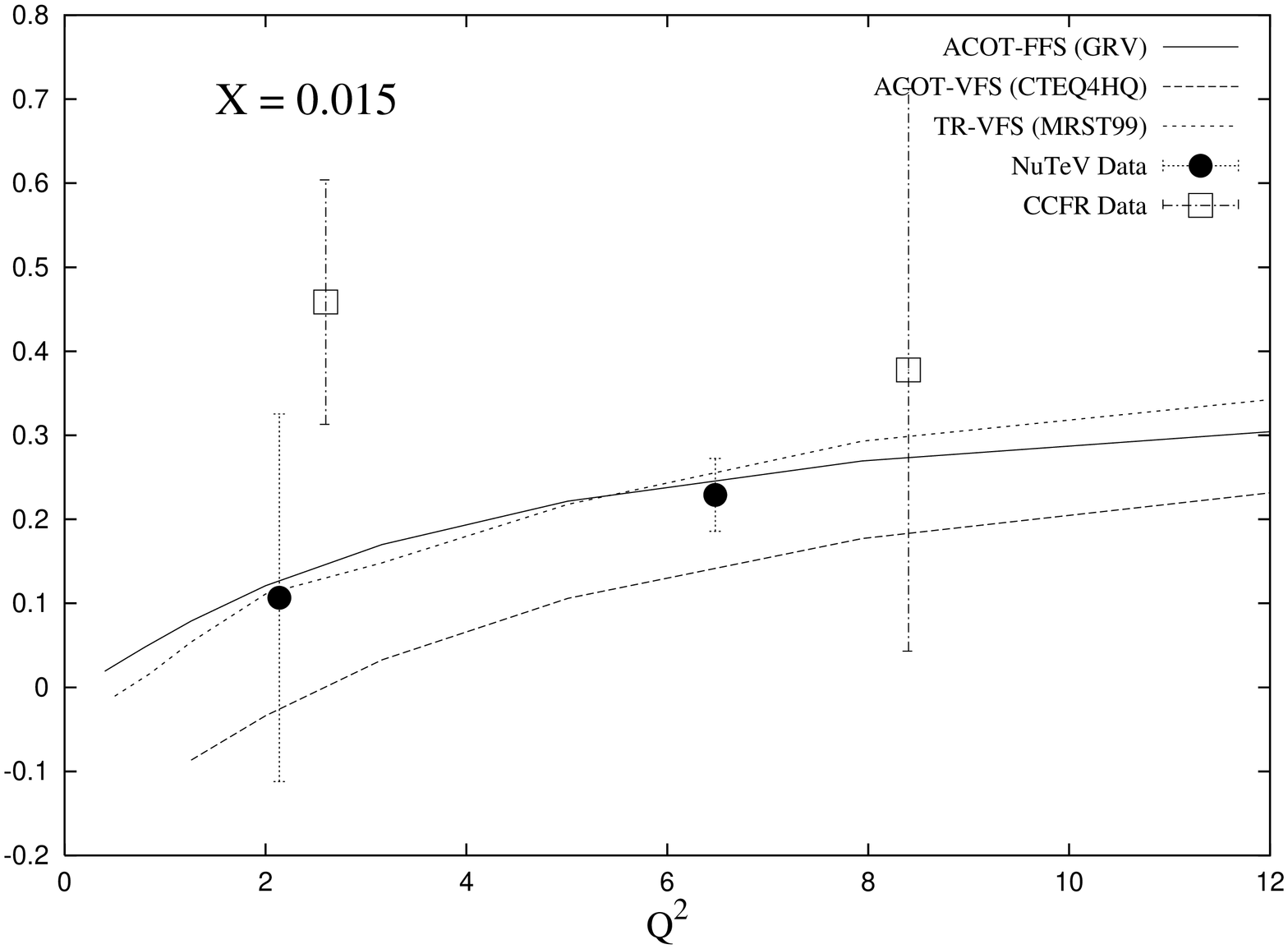,height=2in,angle=270}
\psfig{figure=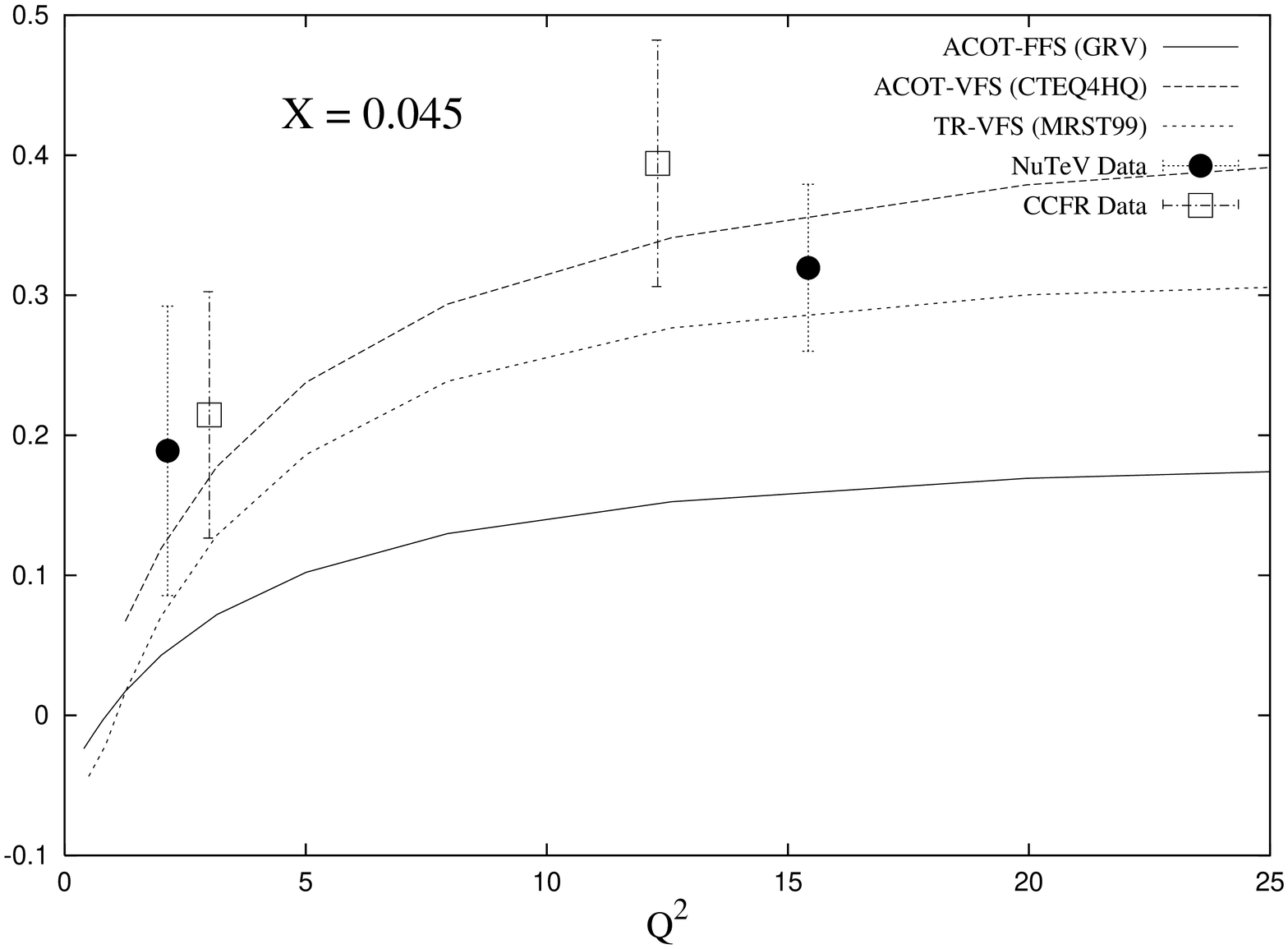,height=2in,angle=270} 
\psfig{figure= 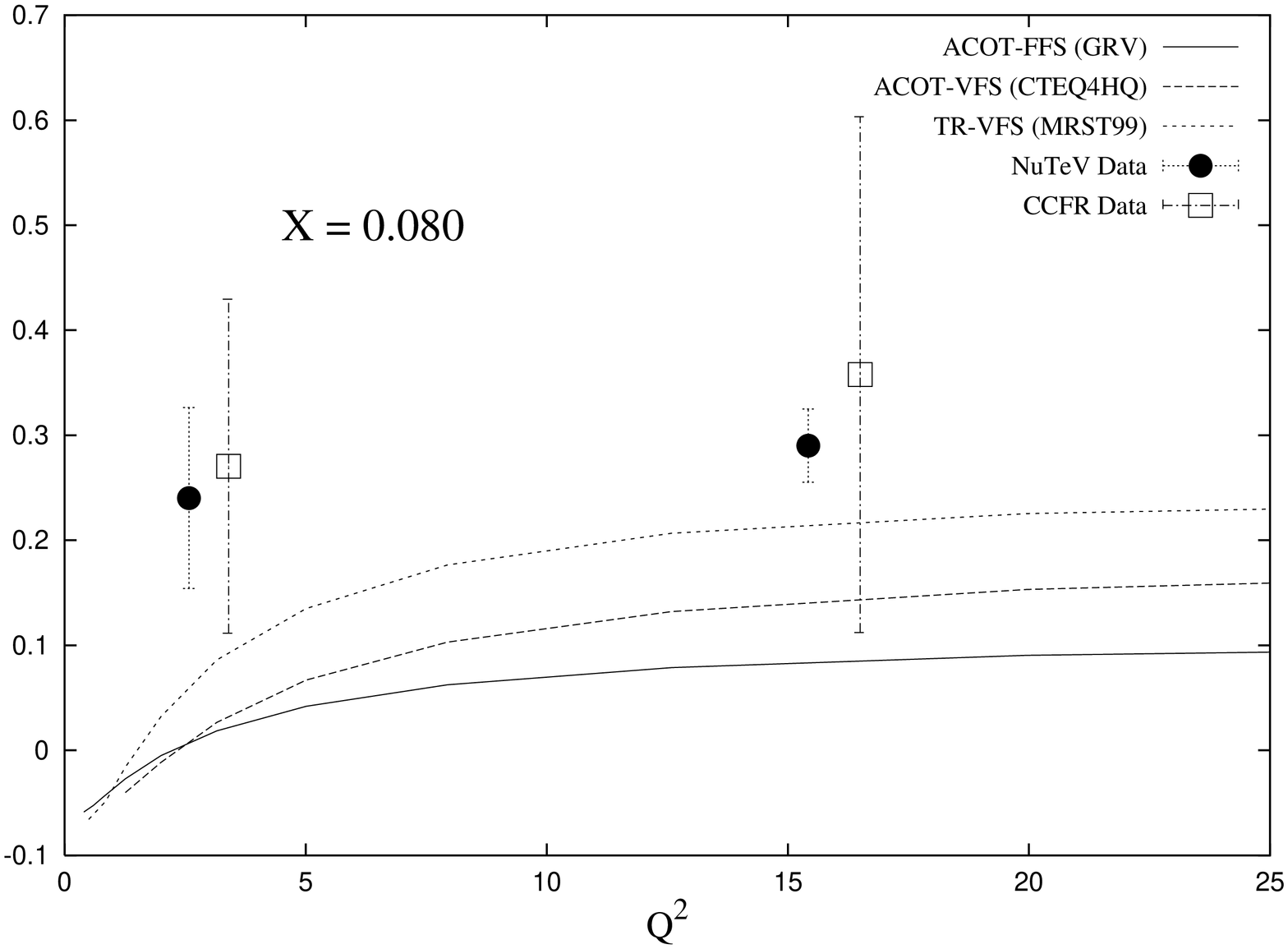,height=2in,angle=270}
\caption{Fit components of $F(\epsilon)$ and model independent
  extraction of $\Delta xF_3$ for CCFR and NuTeV at \mbox{x = 0.015,
    0.045, and 0.080} and NLO models for $\Delta xF_3$ as a function
  of $Q^2$.  NuTeV data are shown as closed circles with statistical
  errors only; CCFR data, open boxes with statistical and systematic
  errors.}
\label{fig:fc}
\end{figure}

A model or extrapolation for either $R(x,Q^2)$ or $\Delta xF_3$ must
be provided as input for any extraction.  The results shown in
Fig.~\ref{fig:fc} use the world parameterization of
$R(x,Q^2)$~\cite{Rworld} and are shown with models of parton
distributions functions from fixed and variable flavor schemes: Shown
are ACOT fixed flavor scheme (ACOT-FFS) using GRV parton
distributions; ACOT variable flavor scheme using CTEQ4HQ (ACOT-VFS);
and Thorne-Roberts variable flavor scheme with MRST 99
(TR-VFS).~\cite{trvfs,acotvfs}

\section*{Conclusions and Prospects}

Preliminary physics results of the NuTeV differential cross section
results have been presented and structure functions have been
extracted.  The results are found to be in good agreement with CCFR
results.

The measurement of $\Delta xF_3$ was performed using a model for
$R(x,Q^2)$.  Due to the positive correlation between $R$ and $\Delta
xF_3$, this measurement depends strongly on the input value for $R$.
The results of this measurement have been found to be in excess of
variable and fixed flavor model schemes, when an extrapolation of the
world's knowledge of $R$ is used to extract $\Delta xF_3$.

\begin{figure}
\mbox{\epsfig{figure=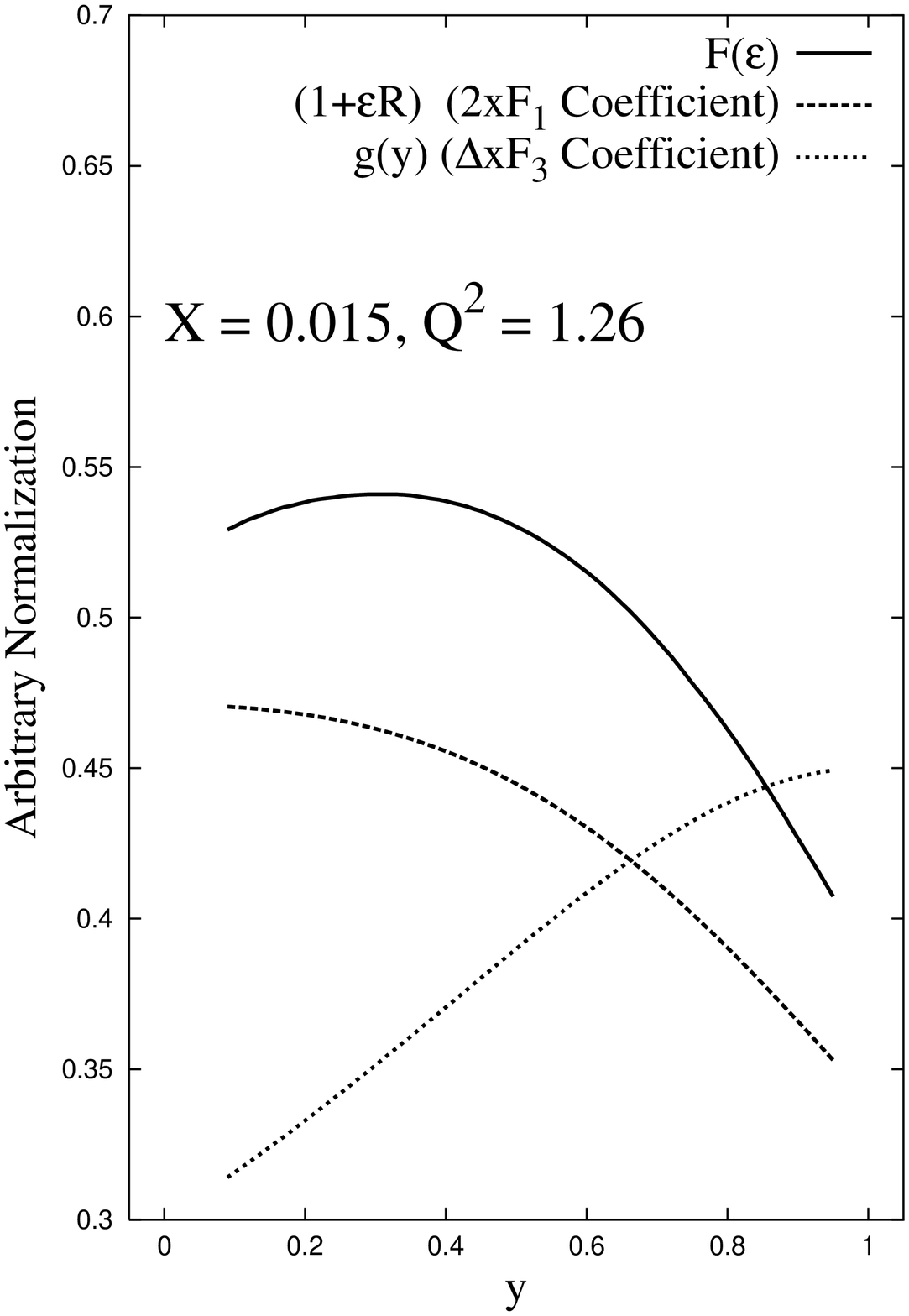,height=3.0in}
\psfig{figure=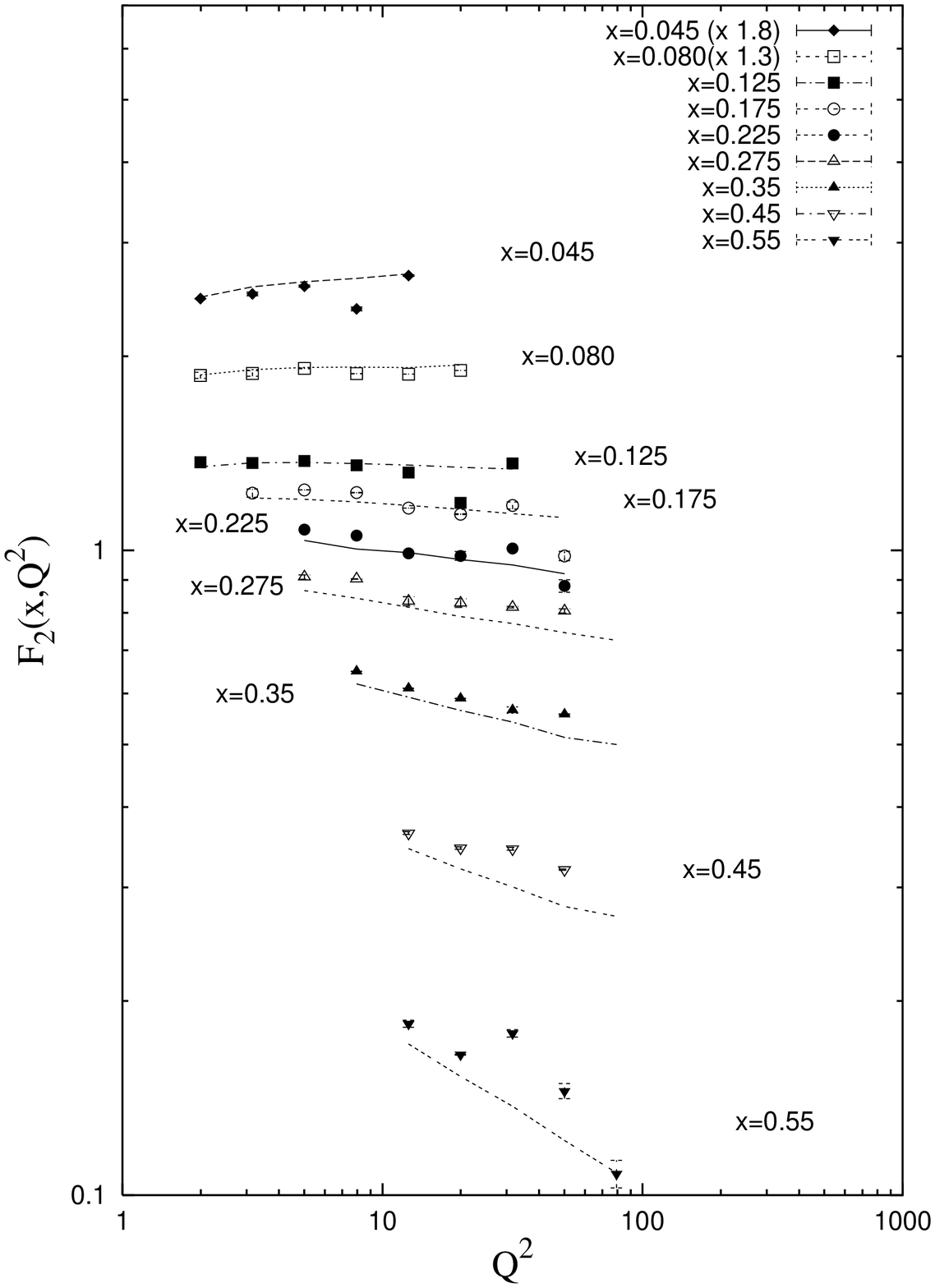,height=3.0in}}
\caption{Left:  An example fit.  The upper curve is
  $F(\epsilon)$ as a function of $y$.  The $y$-intercept effectively
  gives the value of $F_2$, which is largely independent of the choice
  of $R(x,Q^2)$.  Right: The extracted values of $F_2$ from NuTeV and
  a curve showing the theoretical extraction of $F_2$ as a function of
  $Q^2$ with statistical errors.}
\label{fig:f2}
\end{figure}

The prospects for improving upon the current preliminary measurement
and the CCFR measurement rely on the improved calibration for NuTeV
and the extended kinematic range.  The extension of the kinematic
range from NuTeV may provide great insight into the extraction process
(see Fig.~\ref{fig:f2}).  The high-$y$ and low-$y$ data points are the
most sensitive to the underlying structure function quantities.  The
calibration at lower hadronic energies extends the low-$y$ reach to
lower $Q^2$ for moderate $x$.  The sign-selected beam allows the
inclusion of high-$y$ events.  This is where the sensitivity to
$R(x,Q^2)$ and $\Delta xF_3$ is the greatest.

\end{document}